\documentclass[10pt,3p,sort&compress]{elsarticle}
\usepackage{amsmath,amssymb}
\usepackage{epsfig}
\usepackage{graphicx}% Include figure files
\usepackage{bm}
\usepackage{amsmath}
\def\be {\begin{equation}}
\def\ee {\end{equation}}
\def\ba {\begin{eqnarray}}
\def\ea {\end{eqnarray}}
\def\nn {\nonumber}
\newcommand*{\rlo}{R^T}
\newcommand*{\glo}{G^T}

\def\c  {\gamma}

\def\({\left(}
\def\){\right)}
\def\na{\nabla}
\def\[{\left[}
\def\]{\right]}

\def\bi {\begin{itemize}}
\def\ei {\end{itemize}}
\newcommand\beq{\begin{eqnarray}}
\newcommand\eeq{\end{eqnarray}}
\newcommand{\bea}{\begin{eqnarray}}
\newcommand{\eea}{\end{eqnarray}}

\newcommand{\bal}{\begin{align}}
\newcommand{\eal}{\end{align}}

\usepackage{graphicx}% Include figure files
\usepackage{dcolumn}% Align table columns on decimal point
\usepackage{bm,morefloats,color}% bold math

\usepackage{amsmath,latexsym}
\usepackage{amsthm}
\usepackage{tensor}
\usepackage[makeroom]{cancel}

\usepackage{comment}
\usepackage{indentfirst}

\usepackage[
      colorlinks=true,
      linkcolor=blue,
      urlcolor=blue,
      filecolor=blue,
      citecolor=red,
      pdfstartview=FitV,
      pdftitle={},
      pdfauthor={},
      pdfsubject={},
      pdfkeywords={},
      pdfpagemode=None,
      bookmarksopen=true
]{hyperref}

\usepackage{amssymb,bm}

\def\Eqref#1{Eq.~(\ref{#1})}

\def\X5sp{{\rm X}_5}
\def\Y3sp{{\rm Y}_3}
\def\Z3sp{{\rm Z}_3}

\begin{document}

\begin{frontmatter}

\title{
{\bf
Covariant action for bouncing cosmologies in modified\\
Gauss-\hskip -0.06cm Bonnet gravity}}
\author{{{\bf In\^es Terrucha, Daniele Vernieri, Jos\'e P. S. Lemos}}}
\address{\rm Centro de Astrof\'isica e Gravita\c c\~ao {}- CENTRA,
Departamento de F\'{\i}sica, \\Instituto Superior T\'ecnico - IST,
Universidade de Lisboa - UL, \\Avenida Rovisco Pais 1, 1049-001,
Portugal
\vskip 0.13cm
emails: ines.terrucha@tecnico.ulisboa.pt,\\
daniele.vernieri@tecnico.ulisboa.pt, \\joselemos@tecnico.ulisboa.pt
%\ead{ines.terrucha@tecnico.ulisboa.pt,
%daniele.vernieri@tecnico.ulisboa.pt, joselemos@tecnico.ulisboa.pt}
}

\begin{abstract}

Cyclic universes with bouncing solutions are candidates for solving
the big bang initial singularity problem. Here we seek bouncing
solutions in a modified Gauss-\hskip -0.05cm Bonnet gravity theory, of
the type $R+f(G)$, where $R$ is the Ricci scalar, $G$ is the
Gauss-\hskip -0.05cm Bonnet term, and $f$ some function of it.  In
finding such a bouncing solution we resort to a technique that reduces
the order of the differential equations of the $R+f(G)$ theory to
second order equations.  As general relativity is a theory whose
equations are of second order, this order reduction technique enables
one to find solutions which are perturbatively close to general
relativity. We also build the covariant action of the order reduced
theory.

\end{abstract}

%\keywords{}
%\date{\today}

\end{frontmatter}

\section{Introduction}

The physics of the big bang initial singularity is still an open
problem in cosmology and fundamental physics. A complete description
of the universe must avoid at all costs spacetime singularities as
their existence makes the future physically unpredictable.  It is
supposed that in a quantum gravity regime new physics sets in and
spacetimes singularities get a proper description.  In such a regime
there are several possibilities, but an intriguing one is to suppose
that at some tiny scale, of the order of few Planck units, the
universe undergoes a
bounce~\cite{Brandenberger:2016vhg,Chamseddine:2016uef},
such that a previously collapsing
universe expands back again originating our own visible universe, yet
to possibly collapse again in a cyclic way, giving rise to a cyclic
universe~\cite{Steinhardt:2001st}.

A modified Friedmann equation
of the type,
$
H^2=\frac{1}{3}\kappa \rho\(1-\frac{\rho}{\rho_c}\)
$,
has been predicted in 
loop quantum
cosmology~\cite{Bojowald:2001xe,Ashtekar:2006uz,Ashtekar:2006wn,deHaro:2017yll}.
Here,
$H=H(t)$ is the Hubble function defined in terms of the scale
factor $a(t)$ by $H\equiv
\frac{\dot{a}}{a}$, with $t$ being the time
cosmic parameter
and a dot denotes differentiation with respect to $t$,
$\kappa=8\pi$, we put  
Newton's constant ${G}_{{}_N}$ to one
and the speed of light $c$
to one, $\rho=\rho(t)$ is the matter energy density,
and $\rho_c$ is a
critical energy density given by
$\rho_c=\frac{\sqrt{3}}{2\pi\kappa\gamma^3}\rho_{p}$, where
$\rho_{p}$ is the Planck density $\rho_{p}=1$
($\rho_{p}\equiv\frac{1}{{G}_{{}_N}^2\hbar}$, with 
$\hbar$ being the Planck constant that we likewise put equal to one),
$\gamma=\frac{\ln3}{\sqrt8\,\pi}$ is the
Barbero\hskip0.02cm-\hskip-0.02cm{}Immirzi parameter, and
so $\frac{\sqrt{3}}{2\pi\kappa\gamma^3}$ is a number of order $1$.
Thus, in this loop quantum cosmology
scheme the bounce indeed occurs at around the Planck length $l_p$.

It is of interest to reproduce the modified Friedmann equation of the
kind shown through a theory with an appropriate effective covariant
action without having to go into the details of loop quantum
cosmology.  But general relativity gives the pure
Friedmann equation, and the corresponding action, the Einstein-Hilbert
action, is unique in the sense that in four spacetime dimensions it is the only
action that gives a theory with second order field equations for the
metric.  Now, loop quantum gravity, and thus loop quantum cosmology,
has no additional degrees of freedom in relation to general
relativity.  Moreover, in loop quantum cosmology, the modified
Friedmann equation
$
H^2=\frac{1}{3}\kappa \rho\(1-\frac{\rho}{\rho_c}\)
$,
is just the Friedmann equation with an altered source and so clearly
there are no new degrees of freedom. So,  it seems
that without invoking an action with extra degrees of freedom one
cannot get the appropriate modified Friedmann equation obtained
through a quantum version of general relativity, i.e., through loop
quantum cosmology itself.  On the
other hand, invoking an action, and so a theory, with extra degrees of
freedom would generically yield a different modified Friedmann equation.

Sotiriou~\cite{Sotiriou:2008ya} has considered this conundrum in some
detail using $f(R)$ gravity theory~\cite{Sotiriou:2008rp}, where $R$
is the Ricci scalar and $f$ some function of it.  Now, higher-order
theories, such as $f(R)$, go beyond the Einstein-Hilbert action, and
so involve, under metric variation, higher-order differential
equations. These introduce new degrees of freedom such as new extra
gravitational fields.  The upshot is that a modified Friedmann
equation would not look like the one of loop quantum cosmology.
However, by envisaging such a theory as an effective field
theory~\cite{Solomon:2017nlh} and through an order reduction
technique~\cite{Bel:1985zz,Simon:1990ic}, Sotiriou was able to find a
modified Friedmann equation of loop quantum cosmology type without new
degrees of freedom~\cite{Sotiriou:2008ya}.  Thus, even a theory with
new degrees of freedom, as $f(R)$ gravity has in general, can, upon
specific considerations, be treated for some purposes as a theory with
the same degrees of freedom as general
relativity~\cite{Sotiriou:2008ya}.  Such a treatment has also been
used in~\cite{deRham:2017aoj}, but in adopting this procedure one must
be careful, since if order reduction is to be used in conjunction with
effective field theory then one must be aware of the problems it can
raise~\cite{Solomon:2017nlh}.

It is certainly of significance to extend Sotiriou's analysis and see
whether different higher-order gravities can as well yield
bouncing universes.  Here
we consider a modified Gauss-\hskip -0.05cm Bonnet gravity theory with action
$R+f(G)$
~\cite{Nojiri:2005jg,Bamba:2014mya,Haro:2015oqa,Escofet:2015gpa,
Makarenko:2017vuk,Nojiri:2017ncd}, where $G$ is the Gauss-\hskip -0.05cm Bonnet term, given
by $G=R^2-4R^{ab}R_{ab}+R^{abcd}R_{abcd}$, with $R_{abcd}$, $R_{ab}$,
and $R$ being the Riemann tensor, the Ricci tensor, and the Ricci
scalar, respectively, and $f(G)$
means a function of $G$. Although in four spacetime dimensions $G$ is a
topological invariant and does not contribute to the dynamics, $f(G)$
is non trivial even in four dimensions.
In this connection,
it is impressive that modified Gauss-\hskip -0.05cm Bonnet gravity has been
proven to be perfectly viable, since it passes solar system
tests~\cite{Cognola:2006eg} (see also~\cite{Nojiri:2005jg}),
naturally
leads to late-time cosmic acceleration~\cite{Cognola:2006eg},
and is stable under linear
cosmological perturbations~\cite{Bohmer:2009fc}.

Our idea is to have a bouncing cosmology coming out of a still
relatively simple theory such as $R+f(G)$,
while taking into account terms that involve altogether,
the Riemann  and the Ricci
curvatures and  the Ricci scalar.
It would  be interesting too to extend our results for more
general theories including functions of all possible curvature
invariants consistent with the procedure
of effective field theories.

Now, $R+f(G)$ gravity is likewise an higher-order theory and,
as in $f(R)$,
involves under metric variation higher-order differential equations
which in turn introduce new degrees of freedom, and ultimately the
modified Friedmann equation would not be the one we seek, of
the type provided by loop quantum cosmology.  As
in~\cite{Sotiriou:2008ya} one can bypass this problem, sticking to
$R+f(G)$ gravity and using it as an effective field theory of the
higher-order theory together with an order reduction technique
procedure. This approach
avoids the unwanted ghostly modes that arise in higher-order
theories and it further
has the advantage
that the physical solutions one is interested in
can be connected perturbatively to the general relativity solutions.
Some other solutions of the full theory,
like solutions of the 
equivalent scalar-tensor representation of $R+f(G)$ gravity,
may be of interest in other
contexts, but of no use to get the modified Friedmann equation of loop
quantum cosmology.
We note in addition that second-order theories can
still be pathological as, for instance, in the case of a system for
which the equations develop an imaginary propagation speed. We assume
that the order reduction procedure we use does not yield second-order
equations with this unwanted behavior.
One can wonder why the scheme proposed
is a valid
scheme. The answer can lie on the essential features of the fundamental
theory, yet to be formulated, which may impose to be itself
expressed in this way, when
viewed as an effective theory.

In brief, our aim is to find from $R+f(G)$ gravity, an effective
action invariant under diffeomorphisms, that yields a modified field
equation with a modified source that yields a bounce of loop quantum
cosmology type.  We extend thus into $R+f(G)$ gravity the approach and
the results for $f(R)$ theories given in~\cite{Sotiriou:2008ya}.

The paper is organized as follows. In Sec.~\ref{Order reduced f(G)
field equations} we present the $f(G)$ theory and the order reduced
field equations, i.e., the second order field equations that are
perturbatively close to general relativity.  In Sec.~\ref{Modified first
Friedmann equation} we choose the ansatz of a
Friedmann-Lema\^itre-Robertson-Walker (FLRW) line
element and deduce a modified first
Friedmann equation for the evolution of the universe. In
Sec.~\ref{Bouncing cosmology in Gauss-Bonnet modified gravity} we assume
the simplest possible model, i.e., a universe with zero
cosmological constant, zero spatial curvature, and composed of a stiff
fluid and derive the Lagrangian and the conditions obeyed by the
$f(G)$ Gauss-\hskip -0.05cm Bonnet modified gravity to have a universe with a bounce.
In Sec.~\ref{Conclusions} we conclude.

\section{Modified Gauss-\hskip -0.05cm Bonnet  $f(G)$
gravity theory and order reduced field equations}
\label{Order reduced f(G) field equations}

The action for the modified Gauss-\hskip -0.05cm Bonnet $f(G)$ gravity
in four dimensions reads as
\bea
S=S_{\text{grav}}(g_{ab})+
S_{\text{matter}}(g_{ab},\psi)\,,
\label{action0}
\eea
where $S_{\text{grav}}(g_{ab})$ is the appropriate
gravitational action for
the modified Gauss-\hskip -0.05cm Bonnet theory that depends on the metric
$g_{ab}$, 
and $S_{\text{matter}}(g_{ab},\psi)$
is the matter action that depends on the metric
$g_{ab}$ and on matter fields $\psi$.
The gravitational action is
\bea
S_{\text{grav}}=\frac{1}{2\kappa }\int d^4x \sqrt{-g}\,
{\cal L}_{\rm grav}\,,
\label{action}
\eea
where $\kappa=8\pi$,
$g$ is the determinant of the metric $g_{ab}$,
$\small{a,b}$ are spacetime indices, and
${\cal L}_{\rm grav}$ is the Lagrangian density for the
gravity sector given by 
\bea
{\cal L}_{\rm grav}=R+f(G)\,,
\label{action2}
\eea
where $R$ is the Ricci scalar and 
$f(G)$ is a function of the Gauss-\hskip -0.05cm Bonnet term $G$
defined as
\bea
G=R^2-4R^{ab}R_{ab}+R^{abcd}R_{abcd}\,,
\label{g1}
\eea
with
$R_{abcd}$ being the 
Riemann tensor and $R_{ab}$ the Ricci tensor.
In Eq.~(\ref{action2}) the term $R$ is the
Einstein-Hilbert Lagrangian, and the term 
$f(G)$ is the term that yields the modified
gravity we want to consider.

The principle of least action, $\delta S=0$, applied to
Eqs.~(\ref{action0})-(\ref{action2}) yields
\bea
&&R_{ab}-\frac{1}{2}g_{ab}R -\frac{1}{2}g_{ab}f(G)+f'(G)[2RR_{ab}
+2R_{acde}\tensor{R}{_b^c^d^e}-4R_{cb}\tensor{R}{^c_a}-
4\tensor{R}{_a_c_b_d}\tensor{R}{^c^d}] \nonumber \\
&&-4g_{ab}R^{cd}\na_c\na_df'(G)+4\tensor{R}{^c_a}\na_c\na_b f'(G) 
+4\tensor{R}{^c_b}\na_c\na_a f'(G)+2g_{ab}R\Box f'(G)  \nonumber \\
&&-2R\na_a\na_b f'(G)+4\tensor{R}{_a_c_b_d}\na^c\na^d f'(G) 
-4R_{ab}\Box f'(G)=\kappa T_{ab},
\label{eqmotion}
\eea
where $T_{ab}$, defined by $T_{ab}=\frac{-2}{\sqrt{-g}}\frac{\delta
S_{\rm matter}}{\delta g^{ab}}$, is the stress-energy
tensor. In~\Eqref{eqmotion} a prime designates derivation with respect
to $G$, $\na_a$ is the metric covariant derivative and we define the
d'Alembertian as $\Box{}=g^{ab}\na_a\na_b$.
In $f(G)$ gravity, in
contrast to $f(R)$ gravity, the equations of motion~\eqref{eqmotion}
not only depend on $R$ and $R_{ab}$, but also on $R_{abcd}$ and on
$G$ itself. Now, the Riemann tensor can be defined in terms of the Weyl tensor
$C_{abcd}$, the Ricci tensor, the Ricci scalar, and the metric
$g_{ab}$ as
$R_{abcd}=C_{abcd}+
\frac{1}{2}(g_{ac}R_{db}+g_{bd}R_{ca}-
g_{ad}R_{cb}-g_{bc}R_{da}) 
+\frac{1}{6}(g_{ad}g_{cb}-
g_{ac}g_{db})
R$.
We proceed by making 
the simplifying
assumption that the spacetime has zero Weyl tensor,
$C_{abcd}=0$.
This is in line with what we will do next,
when working with a FLRW line element for which the Weyl tensor
$C_{abcd}$ vanishes.
For $C_{abcd}=0$,
we have that $R_{abcd}$ 
can be written in terms of $R_{ab}$ and $R$ as 
\bea
R_{abcd}=
\frac{1}{2}(g_{ac}R_{db}+g_{bd}R_{ca} 
-
g_{ad}R_{cb}-g_{bc}R_{da}) 
+\frac{1}{6}(g_{ad}g_{cb}-
g_{ac}g_{db})R\,.
\label{weyl=0}
\eea
With 
Eq.~(\ref{weyl=0}) one can simplify
Eq.~(\ref{g1}) to
\be
G=\frac{2}{3}R^2-2R_{ab}R^{ab}\,.
\label{g2}
\ee
Equations~(\ref{weyl=0}) and
(\ref{g2})
can then be put directly into
Eq.~(\ref{eqmotion}) defining our
equation of motion.

We may further parametrize the modified
Gauss-\hskip -0.05cm Bonnet function $f(G)$ as
\bea
f(G)=2 \Lambda+\epsilon\varphi(G)\,,
\label{fg1}
\eea
where $\Lambda$ is a cosmological constant term, 
$\varphi(G)$ is a
function of the invariant $G$
and $\epsilon$ is a
dimensionless parameter.

Two comments are in order here
in regard to Eqs.~(\ref{action2}) and
(\ref{fg1}).
First,
one can think of $\varphi$ as a function incorporating all
possible corrections to the Einstein-Hilbert action. For
instance, if $\varphi$ is thought of as an expansion, not
necessarily analytic, then one possibility 
is that it could look like 
$\varphi= \dots + a_1l_p^{-6}/G
+ a_2 l_p^2G + a_3 l_p^2 G
\ln l_p^4G+
a_4  l_p^{6} G^2 +\dots$,
where the characteristic scale was set
to be $l_p$ and
the $a_i$ are coefficients without units.
The expansion could have other different terms, see
~\cite{Nojiri:2005jg,Bamba:2014mya,Haro:2015oqa,
Escofet:2015gpa,Makarenko:2017vuk,Nojiri:2017ncd}
for some other terms that can appear.
Second, the $\epsilon$ parameter
is not a priori a small parameter.
It could, if we wished, be
absorbed in $\varphi$. For instance, in the expansion just given
it could be absorbed by the $a_i$s.
Moreover, clearly
a
parameter such as $\epsilon$ is desired
such that when it goes to zero
general relativity is 
brought back. Indeed, 
when $\epsilon$ is set to zero in Eq.~(\ref{fg1}),
the action~(\ref{action2})
gives simple general
relativity,
so that $\epsilon$
serves essentially to
indicate the deviation from general relativity.

We can now start our plan.
Rather than considering it
as an exact theory, 
we want to consider the $R+f(G)$ gravity, Eqs.~(\ref{action2}) and
(\ref{fg1}),
as an effective field theory
perturbatively close to general relativity in such a way
as to not change the original degrees of freedom.
To do so we, work at lowest
order in $\epsilon$.
For implementing this
plan, from  Eqs.~(\ref{action2}) and~(\ref{fg1})
we see we must have 
that 
$\epsilon \varphi \ll R$\ at the
range of curvatures considered.
For instance, suppose that
$\epsilon\varphi(G)\sim R^2/\rho_c$, for 
some critical density $\rho_c$.
If then the universe's critical density 
at a bounce is of the
order of the Planck density, then
at around that scale
essentially $R \ll \rho_c\sim \rho_p
\sim l_p^{-2}$.

To maintain the original degrees of freedom
of general relativity, and thus get
rid of the unwanted extra degrees of
freedom of the $R+f(G)$ gravity,
we apply an order reduction technique to
the field equations.
The parameter $\epsilon$
allows
to develop in a transparent way
this technique,
working out
to lowest
order Eq.~(\ref{eqmotion}).
To proceed we substitute $f(G)$ and $f'(G)$ with $\epsilon=0$ in
\Eqref{eqmotion}, and express $R$ and $R_{ab}$ at the lowest
order. This process yields, 
\bea
&R^T&=-4\Lambda-\kappa T \label{rt}\\
&R^T_{ab}&=-\frac{\kappa }{2}g_{ab}T-\Lambda g_{ab}+\kappa
T_{ab}\label{riccitmn},
\eea  
where we denote lowest order values by using the superscript $T$.
At lowest order, $\epsilon=0$, and using
Eqs.~(\ref{rt}) and (\ref{riccitmn}),
we  obtain for 
Eqs.~(\ref{weyl=0}) and (\ref{g2}) the following expressions 

\bea
\rlo_{abcd}&=&-\frac{\kappa }{2}(g_{ad}T_{cb}+g_{bc}T_{da} 
-g_{ac}T_{db}-g_{bd}T_{ca}) -\frac{1}{3}(g_{ac}g_{bd} 
-g_{ad}g_{cb})(\Lambda+\kappa T)\,,\\
\glo&=&\frac{2}{3}\kappa^2T^2-2\kappa^2T_{ab}T^{ab}
+\frac{8}{3}\Lambda^2+\frac{4}{3}\Lambda\kappa T\,.
\label{gt}
\eea

The application of order reduction is equivalent to replacing
$R$, $R_{ab}$, $R_{abcd}$, and $G$
by, respectively, 
$R^T$, $R_{ab}^T$, $R_{abcd}^T$, and $G^T$ in \Eqref{eqmotion}.
This procedure brings us to the expression
\bea
&&R_{ab}-\frac{1}{2}g_{ab}R+\epsilon\bigg[-\frac{1}{2}g_{ab}\varphi^T 
+\varphi'^T\bigg(2\rlo\rlo_{ab}-4\rlo_{cb}g^{cd}\rlo_{ad} 
+2\rlo_{acde}g^{cf}g^{dg}g^{eh}\rlo_{bfgh}
-4g^{cd}g^{ef}\rlo_{acbe}\rlo_{df}\bigg)\nn\\
&&
-4\rlo_{ab}\Box\varphi'^T
-4g_{ab}g^{ce}g^{df}\rlo_{ef}\na_c\na_d\varphi'^T 
+4g^{cd}\rlo_{da}\na_c\na_b\varphi'^T+4g^{cd}\rlo_{db}\na_c\na_a\varphi'^T
+2g_{ab}\rlo\Box\varphi'^T
\nn\\
&&
-2\rlo\na_a\na_b\varphi'^T
+4g^{ce}g^{df}\rlo_{aebf}\na_d\na_c\varphi'^T\bigg]=\kappa T_{ab}\,,
\label{oreqmotion}
\eea
where $\varphi^T=\varphi(G^T)$ and $\varphi'^T=\varphi'(G^T)$, with
the prime denoting differentiation with respect to the Gauss-\hskip
-0.05cm Bonnet invariant $G^T$.  Eq.~(\ref{oreqmotion}) is the order
reduced field equation that we wanted.

\section{Modified first Friedmann equation}
\label{Modified first Friedmann equation}
Our next step is to derive a modified first Friedmann equation from
the order reduced field equation, \Eqref{oreqmotion}. To do so we
choose a FLRW line element of the type
\bea
ds^2=-dt^2+a^2\Biggl[\frac{dr^2}{1-kr^2}+r^2(d\theta^2\Biggr. 
\Biggl.+\sin^2\theta d\phi^2)\Biggr]\,,
\label{lineele}
\eea
where $t$ is the time coordinate,
$(r,\theta,\phi)$ are the spatial coordinates,
$a=a(t)$ is the cosmological scale factor,
and $k=-1,0,1$
yields hyperbolic, flat, and spherical spaces, respectively.
We assume  a perfect fluid description for the
fields involved, so the stress-energy
tensor is 
\bea
T_{ab}=\(\rho+p\)u_a u_b+pg_{ab}\,,
\label{perffluid}
\eea
where $u_a$ is the fluid's
4-velocity,
$\rho=\rho(t)$ and $p=p(t)$ are the energy density and pressure of the fluid,
respectively, with $p$ being given by
\bea
p=w\rho\,,
\label{eqstate}
\eea
and $w$ being a number.
Defining the Hubble function $H=H(t)$ by
\bea
H=\frac{\dot{a}}{a}\,,
\label{hf}
\eea
where a dot means derivative with respect to time $t$,
the zero\hskip0.01cm-\hskip-0.01cm zero component of \Eqref{oreqmotion} is 
\bea
\frac{6 k}{a^2}+6 H^2+2 \Lambda -2 \kappa  \rho+\epsilon
\varphi^T 
-\frac{4}{3} \epsilon  (\Lambda -\kappa  \rho)
\left[6 H \dot{\varphi}'^T+\varphi'^T
\big(2 \Lambda \big.\right.
\big.\left.+\kappa\rho(1+3 w)\big)\right]=0\,.
\label{fried1}
\eea
An independent equation can be taken
by noting that, assuming a FLRW metric, i.e., 
Eq.~(\ref{lineele}), one has $\nabla_a T^{ab}=0$
in
Eq.~(\ref{oreqmotion}).
This is equivalent
to energy conservation
which reads
\bea
\dot{\rho}=-3H(1+w)\rho\,.
\label{enercons}
\eea
Now, using the chain rule, one has
$\dot{\varphi}'^T=\frac{\partial \varphi'^T}{\partial \glo}
\frac{\partial \glo}{\partial \rho}
\dot\rho$. Substituting $\dot\rho$ from 
Eq.~(\ref{enercons}) one finds
\bea
\dot{\varphi}'^T&=&4\varphi''^T H \kappa
\rho (1+w)  [\Lambda+2 \kappa \rho (1+3 w) 
-3 \Lambda  w]\,.
\label{chainrule}
\eea
After placing \Eqref{chainrule}  into \Eqref{fried1}, we get
\bea 
&&H^2=-\frac{k}{a^2}-\frac{\Lambda}{3}+\frac{1}{3}\kappa\rho+
\frac{\epsilon}{18}\bigg[-3\varphi^T 
+\frac{96\kappa k\rho\varphi''^T(1+w)(\kappa\rho-\Lambda)(\Lambda
+2\kappa\rho(1+3 w)-3\Lambda w)}{a^2} \nonumber\\
&&+4(\Lambda -\kappa\rho)\bigg(8\kappa\rho\varphi''^T(1+w)
(\kappa\rho-\Lambda)(\Lambda
+2\kappa\rho(1+3 w)-3\Lambda w)
+\varphi'^T(2\Lambda+\kappa\rho(1+3 w))\bigg)\bigg]\,.
\label{friedmannk0}
\eea
When substituting 
Eq.~(\ref{chainrule})
into Eq.~(\ref{fried1}),
we have 
used the lowest order value for $H^2$, namely,
$H^2=-\frac{k}{a^2}-\frac{\Lambda}{3}+\frac{1}{3}\kappa\rho$.
Eq.~(\ref{friedmannk0}) is the modified
first Friedmann equation for
the $R+f(G)$ gravity theory
after a process of order reduction.

\section{Bouncing cosmology in Gauss-\hskip -0.05cm Bonnet modified gravity}
\label{Bouncing cosmology in Gauss-Bonnet modified gravity}

Let us now assume the simplest possible model, i.e., 
zero cosmological constant, zero spatial curvature,
and a stiff fluid so that 
\bea
\Lambda=0\,,\quad k=0\,,\quad w=1\,,
\label{cond}
\eea
respectively.
First, with these assumptions
we have that Eqs.~(\ref{gt}),
(\ref{perffluid}),
(\ref{eqstate}), and 
(\ref{cond})
yield
\bea
\rho^2=-\frac{3}{16\kappa ^2}\glo\,.
\label{rhogt}
\eea
Thus, from Eq.~(\ref{rhogt}) we see that
$\glo$ must be negative.
Then, using \Eqref{rhogt}
and the assumptions of \Eqref{cond},
one finds that  
\Eqref{friedmannk0} turns into
\bea
H^2=\frac13 {\kappa \rho}-\epsilon\left((\glo)^2\varphi''^T
-\frac{1}{6}\glo\varphi'^T\right. 
\left.+\frac{1}{6}\varphi^T\right)\,.
\label{friedmannk01}
\eea
Notice that, by taking $\varphi^T\propto G^T$, 
Eq.~\eqref{friedmannk01}
returns the first Friedmann equation of general
relativity as expected, i.e., $H^2=\frac13 {\kappa \rho}$.

We want to retrieve a cyclic universe dynamical
equation with a bounce. Following e.g.~\cite{Sotiriou:2008ya},
the appropriate Friedmann equation  with a bounce
can be written as
\bea
H^2=\frac{1}{3}\kappa \rho\(1-\frac{\rho}{\rho_c}\)\,,
\label{cyclic}
\eea
where  $\rho_c$ is the critical energy density
at which the bounce occurs.

Comparing Eqs.~(\ref{friedmannk01}) with (\ref{cyclic})
we get
$\epsilon\left((\glo)^2\varphi''^T-\frac{1}{6}\glo\varphi'^T+
\frac{1}{6}\varphi^T\right)
=\frac{\kappa \rho^2}{3\rho_c}$.
This equation upon using 
\Eqref{rhogt} turns into
$
\epsilon\left(-(\glo)^2\varphi''^T+\frac{1}{6}
\glo\varphi'^T -\frac{1}{6}\varphi^T\right)=\frac{G^T}
{16\kappa \rho_c}
$.
In
the solution for $\varphi(G)$ we can drop the superscript $T$
since to $\epsilon$ order $f(G)$ and $f(G^T)$ are the same,
see~\Eqref{fg1}. So, the equation is
\be
\epsilon\left(-G^2\varphi''+\frac{1}{6}
G\varphi' -\frac{1}{6}\varphi\right)=\frac{G}
{16\kappa \rho_c}\,.
\label{difeq0}
\ee
This is a differential equation for
$\varphi$.
Now we are studying the dynamics of the universe
for times that are close to, but still
not quite at the Planck scale,
say 10 times the Planck scale.
Since the quantity $\rho_ c$ sets the scale
of the problem,
we rescale Eq.~(\ref{difeq0})
by writing
$\bar G= G/\rho_c^2$ and $\bar \varphi= \varphi/\rho_c$. So 
Eq.~(\ref{difeq0})
is now
\be
\epsilon\left(-{\bar G}^2{\bar \varphi''}+\frac{1}{6}
{\bar G}{\bar \varphi'} -\frac{1}{6}
{\bar \varphi}\right)=\frac{{\bar G}}
{16\kappa}\,,
\label{difeq}
\ee
where a dash denotes now differentiation with respect to
${\bar G}$.
The solution for \Eqref{difeq} reads 
$
{\bar \varphi(\bar G)}={\bar c}_1\sqrt[6]{{\bar G}}+
{\bar c}_2{\bar G}-\frac{3{\bar G} (5\ln{(-\bar G)}-6)}
{200\kappa \epsilon}
$, where ${\bar c}_1$ and ${\bar c}_2$ are dimensionless
constants of integration. 
Now, $\sqrt[6]{{\bar G}}$ is a term that
is not a correction term in comparison to
$R$, so we put ${\bar c}_1=0$
and the ${\bar G}$ is a pure divergence
so we can drop the corresponding terms. So
finally 
we get the solution
$
{\bar \varphi(\bar G)}=
\frac{3}{40\kappa \epsilon}{\lvert\bar G\rvert} \ln{\lvert\bar G\rvert}
$, where $\lvert\bar G\rvert$ means the absolute value of
$\bar G$.
Recovering $\varphi$ in terms of $G$ we have
\be
\epsilon\varphi(G)=
\frac{3}{40\kappa }  \frac{\lvert G\rvert}{\rho_c} \ln{
\frac{\lvert G\rvert}{\rho_c^2}}\,.
\label{soldifeq}
\ee
Then, the Lagrangian for the gravity sector of the modified
Gauss-\hskip -0.05cm Bonnet $R+f(G)$ gravity is
$\mathcal{L}_{\rm grav}=R+f(G)=R+\epsilon\varphi(G)$, see~\Eqref{action2},
which after using   
\Eqref{soldifeq} gives 
\bea
\mathcal{L}_{\rm grav}=R+\frac{3}{40\kappa }
\frac{\lvert G\rvert}{\rho_c} \ln{ \frac{\lvert G\rvert}{\rho_c^2}}\,.
\label{soldifeq2}
\eea
This is the Lagrangian we were after.

The approximation we use is valid for $R\ll\rho_c\sim l_p^{-2}$.
So,
at $\rho_c$ the approximation breaks down,
since there $R\sim\rho_c\sim l_p^{-2}$ and
one cannot deduce that~\Eqref{soldifeq2}
yields~\Eqref{cyclic} at the bounce.
Our claim is that at stages close to the
bounce~\Eqref{cyclic} holds still as a good approximation
to our Lagrangian, describing
correctly 
the collapsing universe just before and just after
the bounce. Typically, 
the equation is valid for $\rho\sim0.1\rho_c$ say,
in which case a typical length $l$ would
have $l\sim 3l_p$. On the other
hand~\Eqref{cyclic} is an equation that
comes from loop quantum gravity and
so supposedly correct  at the
bounce itself. Thus, presumably, the Lagrangian~\Eqref{soldifeq2}
holds good
for the bounce.

We  now note that the Lagrangian for the $f(R)$ gravity theory yields
that the first order correction to $R$ is an $R^2$ term, indeed
$\mathcal{L}_{ {\rm grav}{f(R)}}=R+\frac{1}{18\kappa}
\frac{R^2}{\rho_c}$~\cite{Sotiriou:2008ya}.
This Lagrangian is to be compared to the Lagrangian we
have found in~\Eqref{soldifeq2} for the $R+f(G)$ theory.
\Eqref{soldifeq2} has the zeroth order Einstein-Hilbert term $R$ plus
the term $\lvert G\rvert\ln \lvert G\rvert$ as the first order
correction to general relativity.

Interesting to note that
Lagrangians of the type of our Eq.~(\ref{soldifeq2}) were  found too
using different approaches where some reconstruction methods were used
by demanding that the scale factor undergoes a bounce at some
cosmological time~\cite{Bamba:2014mya,Haro:2015oqa,Escofet:2015gpa,
Makarenko:2017vuk,Nojiri:2017ncd}. In this way specific $R+f(G)$
gravity bouncing models in the early universe were  selected.

\section{Conclusions}
\label{Conclusions}

We have derived an effective Lagrangian, and so a covariant action,
for the modified Gauss-\hskip -0.05cm Bonnet $R+f(G)$ gravity theory
which yields a cosmological solution for a bouncing universe.  Since
the theory under study generically involves higher-order differential
equations which might lead to physical instabilities, we have used an
order reduction technique procedure. By means of this approach, one
can indeed avoid the unwanted ghostly modes arising in these theories
to construct viable physical solutions which can be connected
perturbatively to the solutions of general relativity. Notice that
higher-order derivatives are not the only source of possible
instabilities since also second order theories can lead to
pathologies. However, in order to give a de\-fi\-ni\-te answer about
this issue, one would need to make a thorough perturbative analysis
such as to understand if the proposed bounce solution is smooth and
dynamically viable.

Let us in addition stress that there has been a
growing interest in the understanding of the big bang initial
singularity, and universes with a bounce of the sort we have presented
here are viable candidates to finding a solution to this problem.
It would be interesting to investigate, by means of the
same approach used here, more general theories including functions of
all possible curvature invariants.

\section*{Acknowledgments}
We acknowledge financial support from the FCT Project
No.~UID/FIS/00099/2013.  DV acknowledges financial support through the
FCT Project IF/00250/2013.  JPSL acknowledges the FCT Grant
No.~SFRH/BSAB/128455/2017 and the Grant No.~88887.068694/2014-00 from
Aperfei\c coamento do Pessoal de N\'\i vel Superior (CAPES), Brazil.
JPSL is grateful to the Gravitational Physics Group at the
University of Vienna for hospitality.

\vskip 1cm
%\newpage

%%%%%%%%%%%%%%%%%%%%%%%%%%%%%%%%%%%%%%%%%%%%%%%

%\noindent {\bf References}
%\vskip 0.2cm

\end{document}